\documentclass[twocolumn,a4paper,showpacs,preprintnumbers,amsmath,amssymb,nofootinbib,floatfix]{revtex4}
\def \be {\begin{equation}}

\def \ee {\end{equation}}
\def \bea {\begin{eqnarray}}
\def \eea {\end{eqnarray}}
\usepackage{natbib}
\usepackage{bm}
\usepackage{graphicx}% Include figure files
\usepackage{dcolumn}% Align table columns on decimal point
\usepackage{bm}% bold math
\usepackage{epsfig}

\begin{document} 

\title{Gravitational lens time-delay as a probe of a possible time variation of the fine-structure constant}

\author{L. R. Cola\c{c}o$^{1}$} \email{colacolrc@gmail.com} 
\author{J. E. Gonzalez$^{2,1}$} \email{gonzalezsjavier@gmail.com}
\author{R. F. L. Holanda$^{1}$} \email{holandarfl@gmail.com}
%\author{F. S. Lima$^{2}$} \email{limasdl@bol.com.br}
\affiliation{$^{1}$Universidade Federal do Rio Grande do Norte, Departamento de F\'{\i}sica, Natal - Rio Grande do Norte, 59072-970, Brasil}
\affiliation{$^{2}$Facultad de Ciencias e Ingeniería, Universidad de Manizales, 170002, Manizales, Colombia}

\begin{abstract}

A new  method based on large scale structure observations is proposed to probe a possible temporal variation of the fine-structure constant ($\alpha$). Our analyses are based on time-delay of Strong Gravitational Lensing and Type Ia Supernovae observations. By considering the runaway dilaton scenario, where the cosmological temporal evolution of the fine-structure constant is given by $\frac{\Delta \alpha}{\alpha} \approx -\gamma \ln{(1+z)}$, we obtain  limits on the physical properties parameter of the model ($\gamma$) at the level $10^{-2}$ ($1\sigma$). Although our limits are less restrictive than those obtained by quasar spectroscopy, the approach presented here provides new bounds on the possibility of  $\frac{\Delta \alpha}{\alpha} \neq 0$ at a different range of redshifts.

\end{abstract}
\pacs{98.80.-k, 95.36.+x, 98.80.Es}
\maketitle

\section{Introduction}

It is well-known that the standard physics is characterized by a set of  laws and fundamental couplings which were historically assumed to be space-time invariant. One of the first contributors to ask about this conjecture was  Dirac (1934) \cite{Dirac:1938mt},  arguing that fundamental couplings might not be pure numbers that occur in many theories, but {  they might depend on the state of the Universe}. Thereafter, many theoretical and observational approaches have come searching for a space-time variation of the fundamental couplings of nature (see a detailed review in \cite{Uzan:2010pm,Martins:2017yxk}). Although the search for a possible varying fundamental couplings has raised the interest of many cosmologists, the general relativity theory prohibits any violation since it would violate the Equivalence Principle \cite{Ray:2019lxv}.

In most extensions of the current standard cosmological model, the fundamental couplings are expected to vary leading to  {  consequences that need to be probed} with observational data \cite{Bekenstein256,SBM,BL,BG,MWMK,BBCMAA,Chodos,wuwang,Kiritsis}.  In the astronomical context, particularly from white dwarfs astronomical observations, constraints on $\Delta \alpha/\alpha$ at the level $(2.7 \pm 9.1) \times 10^{-5}$ were obtained by using gravitational potential \cite{Landau:2020vkr,Bainbridge:2017lsj}, where $\alpha$ is the fine structure constant\footnote{$\alpha \equiv e^2/\hbar c$, where $e$ is the elementary charge, $\hbar$ the reduced Planck's constant, and $c$ the speed of the light.}. More recently, 4 new spectral observations of very high redshift quasars,  up to $z \approx 7.1$, have shown no evidence for a temporal variation of the fine-structure constant (by the so-called many-multiplet method) \cite{Wilczynska}. However, when the authors combined those measurements with a large existing sample at lower redshifts, {  it was pointed out that a spatial variation of $\alpha$ is preferred over a no-variation model at the $3.7 \sigma$ level} (see other discussions about $\alpha$ spatial variation in \cite{webb1999,Ubachs:2017zmg}). {  However, very recently, the authors of the Ref. \cite{Lee:2021kjr} showed that fitting turbulent models in quasars necessarily generates or enhances model non-uniqueness, adding a substantial additional random uncertainty to $\Delta \alpha /  \alpha$. In other words, there is a degeneracy between the absorption structure and turbulent models, each giving different $\Delta \alpha / \alpha$ values.}

By using the physics of cosmic microwave background radiation (CMB), the Ref. \cite{Hart:2019dxi} presented updated constraints on the variation of the fine-structure constant  and the effective electron rest mass $m_e$ during the cosmological recombination era. The authors showed that $\alpha$ and $m_e$ can straightly modify the {  recombination} history at $z \approx 1100$, and thus change the temperature and polarization anisotropies of the CMB measured meticulously with the Planck satellite. Although the constraints on $\alpha$ are slightly tightened due to improved Planck 2018 Polarization data \cite{Aghanim:2019ame,Aghanim:2018oex}, the new results remain very similar in relation to the previous CMB analyses \cite{Ade:2014zfo} (see \cite{Smith:2018rnu} for spatial variation of $\alpha$ using CMB data). It is important to emphasize that analyses using CMB data rely on the assumptions of an almost scale-invariant power spectrum and purely adiabatic initial conditions without primordial gravity waves, so that the CMB constraints on varying constants are only competitive for very specific classes of models that predict strong variations in the very early universe. There are other probes using distinct astrophysical observable, such as black hole in a high gravitational potential \cite{Hees:2020gda}, galaxy cluster \cite{galli}, Big-Bang Nucleosynthesis \cite{NBB}, among others \cite{Milakovic:2020tvq,Kraiselburd:2018uac}. Nonetheless, in the Ref. \cite{ZhangGengYin} it was revisited the so called $\Lambda (\alpha)$CDM framework where the cosmological constant is $\Lambda \propto \alpha^{-6}$. By using cosmological observations as SNe Ia, BAO and CMB along with 313 data from absorption systems in the spectra of distant quasars, constraints on two specific $\Lambda (\alpha)$CDM models were performed.{  The authors found $ \frac{\Delta \alpha}{\alpha} \approx 10^{-4}$,} very similar to the results discussed by \cite{Wei2017}.  On the other hand, variations in $\alpha$ have also been explored on the Earth with atomic clock measurements \cite{Hinkley2013} and  isotope ratio measurements \cite{Dijck:2020kfb}, where its sensitivity (around $10^{-18}$) provides a useful constraint on a possible temporal variation of $\alpha$. 

Among  scenarios beyond standard model {  that produce a temporal variation of alpha}, we can cite a particular class of string theory inspired-models\footnote{String theories at low energy predict the existence of dilaton, a scalar partner of Spin-2 graviton.}, the so-called runaway dilaton model 
\cite{damour1,damour2}. In this scenario, the runaway of the scalar field dilaton towards strong\footnote{In addition, this scenario provides a way to reconcile a massless dilaton with experimental data.} coupling may yield a temporal variation of the fine-structure constant. In this context, a possible  evolution of $\alpha$ at low and intermediate redshifts is given by $\frac{\Delta \alpha}{\alpha} \approx -\frac{1}{40}\beta_{\mathbf{had},0}\phi_{0}^{'} \ln{(1+z)} = -\gamma \ln{(1+z)}$, where $\gamma \equiv \frac{1}{40}\beta_{\mathbf{had},0}\phi_{0}^{'}$, $\beta_{\mathbf{had},0}$ is the current value of the coupling between the dilaton and hadronic matter, and $\phi_{0}^{'} = \frac{\partial \phi}{\partial \ln{a}}$ at the present time.

In order to check a possible temporal variation of the fine-structure constant in runaway dilaton scenarios, some methods using astronomical data have been developed in recent years. By using galaxy clusters observations, for instance, the Ref. \cite{colaco2017} proposed an approach by using gas mass fraction (GMF) measurements and luminosity distances of type Ia supernovae (SNe Ia) to put constraints on $\Delta \alpha/\alpha$. The GMF measurements used in the analyses were obtained via the Sunyaev-Zeldovich (SZ) effect at 148 GHz by the Atacama Cosmology Telescope, and the SNe Ia data from the Union2.1 compilation. The results showed no strong evidence for $\Delta \alpha /\alpha\neq 0$. More recently, the Ref.  \cite{colaco2019} argued that the galaxy cluster scaling-relation $Y_{SZ}D_{A}^{2}/C_{XSZ}Y_X$ can also be used to put constraints on the runaway dilaton model. The authors found that $Y_{SZ}D_{A}^{2}/C_{XSZ}Y_X \propto \alpha^{3}$ by considering a direct relation between a temporal variation of the fine-structure constant and a possible deviation of the cosmic distance duality relation (see also \cite{kbora}). Once again, the results showed no strong evidence for $\Delta \alpha /\alpha\neq 0$. Several other tests capable of probing such temporal variation of $\alpha$ with galaxy cluster observations have been emerging since then (see e.g. \cite{colaco2020,holanda2016.1,holanda2016.2} and references therein).

{  In this paper, we present a new method based on time-delay of strong gravitational lensing (SGL) systems and type Ia of supernovae observations to obtain  limits on a possible temporal variation of the fine-structure constant in runaway dilaton scenario. The samples used to perform our approach are: 19 two-image time-delay lensing systems compiled by the Refs. \cite{ibpsc,holicow} jointly with  $1048$ spectroscopically confirmed SNe Ia compiled by \cite{pantheon}. Moreover, we consider a specific catalog containing $158$ confirmed sources of strong gravitational lensing systems from the Ref.\cite{Leaf2018lfu}. We obtain limits on the physical properties parameter of the runaway dilaton model ($\gamma$) at the level $10^{-2}$ ($1\sigma$) in full agreement with recent limits by using galaxy clusters observations plus SNe Ia observations.}

The work is organized as follows: in Section II we develop our methodology. {  The theoretical framework is discussed in Section III}. In Section IV we present the data set used to perform the analyses. In Section V the corresponding statistical analyses and discussions, and in Section VI we finished with the conclusions.

\section{Methodology}

\subsection{Strong Gravitational Lensing Systems}

Strong Gravitational Lensing Systems (SGL) can be used to investigate gravitational and cosmological theories and fundamental physics. Particularly, observed SGL systems and detected by SLACS, LSD, SLS2, and BELLS surveys have been largely used to fit observational bounds on different cosmological parameters.{  An useful quantity in this context should be Einstein ring ($\theta_E$)}. Under the assumption of the singular isothermal sphere (SIS) model to describe lens mass distribution, the Einstein radius $\theta_E$ is given by \cite{sef,Refsdal,Cao2015}:

\begin{equation}
    \theta_E = \frac{4\pi \sigma_{SIS}^{2}}{c_s^2}\frac{D_{A_{ls}}}{D_{A_s}},
    \label{thetaE}
\end{equation}
where $D_{A_{ls}}$ is the angular diameter distance (ADD) from the lens ($l$) to the source ($s$), $D_{A_{s}}$  the ADD from the observer to the source,{$c_s$ is the speed of light between  source and observer}, and $\sigma_{SIS}$ is the velocity dispersion via SIS model. From Eq. \ref{thetaE} the multiple-image separation of the source depends only on the lens and source angular diameter distances. Nonetheless, the quantity of interest is

\begin{equation}
    D \equiv \frac{D_{A_{ls}}}{D_{A_s}} = \frac{\theta_Ec_s^2}{4\pi\sigma_{SIS}^{2}},
\end{equation}
which can be written in terms of the fine-structure constant ($\alpha_s \equiv e^2/\hbar c_s $) by:

\begin{equation}
D=\frac{\theta_Ee^4}{4\pi \alpha_s^2\hbar^2\sigma_{SIS}^{2}}.
\end{equation}
In order to obtain D from SGL systems observations one needs to make an assumption on the variation of alpha. Currently available data make the assumption $\alpha_s = \alpha_0$, the local value of the fine structure constant.\footnote{ In type of theory explored in the present work (see section III), that predicts variation of the fine-structure constant, such variation can arise either from a varying $\mu_0$ theory (vacuum permeability) or from a varying charge of the elementary particles theory. Both interpretations lead to the same modified expression of the fine-structure constant \cite{Uzan:2010pm,hees,Observables,Bekenstein256}.}

\subsection{Time-Delay Systems}

Time-Delay is another important observational consequence of Strong Gravitational Lensing and it can also be used as a powerful astrophysical tool. Based on the fact that photons follow null geodesics and they are originated from a distant source with distinct optical paths, they shall pass through dissimilar gravitational potentials \cite{Cao2015,SHSUYU,TTREU}. Thus, the time-delay is caused by the difference in length of the optical paths and by the gravitational temporal variation originated in the passage through the effective gravitational potential of the lens. 

Time-delay gives a correlation among the angular diameter distances from observer to lens $(D_{A_{l}})$, from observer to source $(D_{A_{s}})$, and from lens to source $(D_{A_{ls}})$ by \cite{Suyu}:

\begin{equation}
    \Delta \tau = \frac{(1+z_l)}{c_s}\frac{D_{A_l}D_{A_s}}{D_{A_{ls}}} \left[  \frac{1}{2}(\vec{\theta}-\vec{\beta})^2 - \Psi(\vec{\theta})   \right],
\end{equation}
where $\Delta \tau$ is the so-called time-delay, $\vec{\theta}$ and $\vec{\beta}$ are, respectively, the angular positions of the image and the source, $z_l$ the lens redshift, and $\Psi$ is  the lens effective gravitational potential. Thereafter, for a two image lens system ($A$ and $B$) with SIS mass profile describing the lens mass, we can obtain \cite{JLWEI}:

\begin{equation}
    \Delta t = \Delta \tau (A) - \Delta \tau (B) = \frac{(1+z_l)}{2c_s}\frac{D_{A_l}D_{A_s}}{D_{A_{ls}}} [\theta_{A}^{2} - \theta_{B}^{2}].
\end{equation}
Defining the quantity $\frac{D_{A_l}D_{A_s}}{D_{A_{ls}}}$ as time-delay angular diameter distance $D_{A_{\Delta t}}$:

\begin{equation}
    D_{A_{\Delta t}} \equiv \frac{D_{A_{l}}D_{A_{s}}}{D_{A_{ls}}} = \frac{2c_s\Delta t}{(1+z_l)(\theta_{A}^{2}-\theta_{B}^{2})},
    \label{Datdef}
\end{equation}
which can be rewritten in terms of $\alpha_s$ by:

\begin{equation}
    D_{A_{\Delta t}} = \frac{2e^2\Delta t}{\hbar \alpha_s(1+z_l)(\theta_{A}^{2}-\theta_{B}^{2})}.
\end{equation}
Also in this case one needs to make an assumption on the variation of alpha to obtain $D_{A\Delta t}$ from the measurements of time delay, and currently available data set $\alpha_s = \alpha_0$.

\section{Theoretical Framework}

\subsection{Scalar-Tensor Theory of Gravity}

Theories of modified gravity associated to a
scalar field with a non-minimal multiplicative coupling to
the usual electromagnetic Lagrangian lead to violations of the Einstein Equivalence Principle (EEP) in the electromagnetic sector. The matter Lagrangian of this type of theories is given by \cite{hees,Minazzoli}

\begin{equation}
    S_{\mathbf{mat}} = \sum_i \int d^4x \sqrt{-g} h_i(\phi) \mathcal{L}_i(g_{\mu \nu}, \Psi_i),
\end{equation}
where $h_i(\phi)$ is a function of the scalar field $\phi$, and $L_i$ are the Lagrangians for the different matter fields. In this context, the fine-structure constant and the cosmic distance duality relation (CDDR) change with cosmological time, and both are intimately and unequivocally related to each other by \cite{hees,Minazzoli}:

\begin{equation}
    \frac{\Delta \alpha}{\alpha}(z) =\frac{h(\phi_0)}{h(\phi)}-1 = \eta^{2}(z)-1,
\end{equation}
where $\eta$ takes into account any deviations of the CDDR. Considering $\alpha = \alpha_0\phi(z)$, where $\alpha_0$ is the current value of $\alpha$, the equation (9) gives $\eta^2(z) = \phi(z)$ \cite{hees}. Therefore, the equations (3) and (7) shall be rewritten, respectively, by

\begin{equation}
    D = D_0\phi^{-2}(z_s)
    \label{D1}
\end{equation}
and
\begin{equation}
     D_{A_{\Delta t}} = D_{A_{\Delta t,0}}\phi^{-1}(z_s),
     \label{Dat1}
 \end{equation}
where, $D_0 \equiv \frac{e^4\theta_E}{4\pi\alpha_{0}^{2}\hbar^{2}\sigma_{SIS}^{2}}$ and $D_{A_{\Delta t,0}} \equiv \frac{2e^2\Delta t}{\hbar\alpha_0(1+z_l)(\theta_{A}^{2}-\theta_{B}^{2})}$. {   It is important to clarify that the subscript $0$ denotes the current quantities that have been obtained in literature by assuming $\alpha_s=\alpha_0$. It does not mean local values of $D$ or $D_{A_{\Delta t}}.$ The $D$ or $D_{A_{\Delta t}}$  quantities (Eq.(10) and Eq.(11)) are the true observational values if one corrects a possible $\alpha$ variation}.

By considering the definitions $D \equiv \frac{D_{A_{ls}}}{D_{A_s}}$ and $D_{A_{\Delta t}} \equiv \frac{D_{A_{l}}D_{A_{s}}}{D_{A_{ls}}}$ jointly with the Eq. (\ref{D1}) and  (\ref{Dat1}), it is possible to obtain:

\begin{eqnarray}
  D_{A_l}&=& D_0 D_{A_{\Delta t,0}} \phi^{-3}(z_s) \nonumber \\ 
   \Longrightarrow && \phi^{3}(z_s) = \frac{D_0D_{A_{\Delta t,0}}}{D_{A_l}}.
\end{eqnarray}

Finally, to perform our tests and impose new limits on a possible time-variation of $\alpha$, it is necessary to know $D_{A_l}$ for each SGL. This quantity can be obtained by SNe Ia luminosity distance measurements with identical redshifts from those of the SGL system sample. {  Since the SNe Ia observations also can are affected by a varying $\alpha$, we shall consider a deformed CDDR as $D_{A_l} = \eta^{-1}(z_l)(1 + z_l)^{-2}D_{L_l}$ \cite{bxuqhuang,cmapscorasaniti} and the fact that $\eta^2(z) = \phi (z)$ \cite{hees}. Then, one may obtain:}

\begin{equation}
    \frac{\phi^{1/2}(z_l)}{\phi^{3}(z_s)} = \frac{D_{L_l}}{D_0 D_{A_{\Delta t,0}}(1+z_l)^2}.
\end{equation}
This is our key equation, if $\alpha=\alpha_0$ the right side is equal to unity and no time variation of $\alpha$ is possible. It is possible to use this expression to impose new limits on a possible time-variation of $\alpha$ in the context of a very specific string-inspired model, the so-called runaway dilaton Model. 

\subsection{Runaway Dilaton Model}

 In this paper, we focus on the runaway dilaton model \cite{damour1,damour2}. The main idea behind such model is exploiting the string-loop modification of the four dimensional effective low-energy action, where its Lagrangian is given by:

\begin{equation}
    \mathcal{L} = \frac{R}{16\pi G}-\frac{1}{8\pi G}(\nabla \phi)^2-\frac{1}{4}B_F(\phi)F^2 + ...,
\end{equation}
where $R$ is the Ricci scalar, and $B_F(\phi)$ is the gauge coupling function.{  The Runaway Dilaton model is a particular case of scalar-tensor theories of gravity commented in section III \cite{hees,damour1,damour2}}. One can show that the Friedmann equation in this scenario is as follows

\begin{equation}
    3H^2 = 8\pi G\sum_{i} \rho_i + H^2\phi'^2,
\end{equation}
where the sum is over the components of the universe, and $H$ is the Hubble parameter. The relevant parameter of this model is the coupling of $\phi$ to hadronic matter. Nevertheless, the runaway of the dilaton towards {  strong coupling can lead to temporal variations} of $\alpha$, and its variation at low and intermediate redshifts is given by  \cite{Martins:2017yxk}

\begin{equation}
    \frac{\Delta \alpha}{\alpha} \approx -\frac{1}{40}\beta_{\mathbf{had},0}\phi_{0}^{'}\ln{(1+z)},
\end{equation}
where $\beta_{\mathbf{had},0}$ is the current value of the coupling between the dilaton and hadronic matter, and $\phi_{0}^{'} \equiv \frac{\partial \phi}{\partial \ln{a}}$.

\section{Reconstruction method and data}

\subsection{Gaussian Processes}
{In order to obtain a continuous regression of the luminosity distance and $D_0$ in function of $z$, we apply the Gaussian Processes (GP) method to the  data sets described in Secs. \ref{SnIa} and \ref{ER}. The GP reconstruction is performed by choosing a prior mean function  and a covariance function which   quantifies the correlation between the values of the dependent variable of the reconstruction and is characterized by a set of hyperparameters. In our reconstructions of $D_0$ and $D_L$, we choose  zero  as the prior mean function to avoid biased results and a Gaussian kernel as covariance function given by:}

\begin{equation}
k(z,z')=\sigma^2 \exp\left(-\frac{(z-z')^2}{2l^2}\right),
\end{equation}{where $\sigma$ and $l$ are hyperparameteres related to the variation of the estimated function and its smoothing scale, respectively. To optimize the hyperparameter values, we maximize the logarithm of the marginal likelihood}\footnote{This expression assumes that the prior mean function is equal to 0 and we omit the term that depends on the number of data points.}:

\begin{equation}
\ln {\mathcal{L}}=-\frac{1}{2}\bm y^T[k(\bm x,\bm x)+C]\bm y-\frac{1}{2}\ln |k(\bm x,\bm x)+C|,
\end{equation}
{where  $\bm{ x}$ and $\bm y$ are the vectors of the independent and dependent data variables, respectively, and C is the covariance matrix of the data (error matrix). We use the  code GaPP\footnote{https://github.com/carlosandrepaes/GaPP} to perform the GP reconstruction of the $D_L$ and $D_0$ data (see Ref. \cite{Seikel:2013fda} for more details about GP).}

\subsection{Supernova Type Ia}
\label{SnIa}
\begin{figure}[htb!]
\centering	
\includegraphics[scale=0.55]{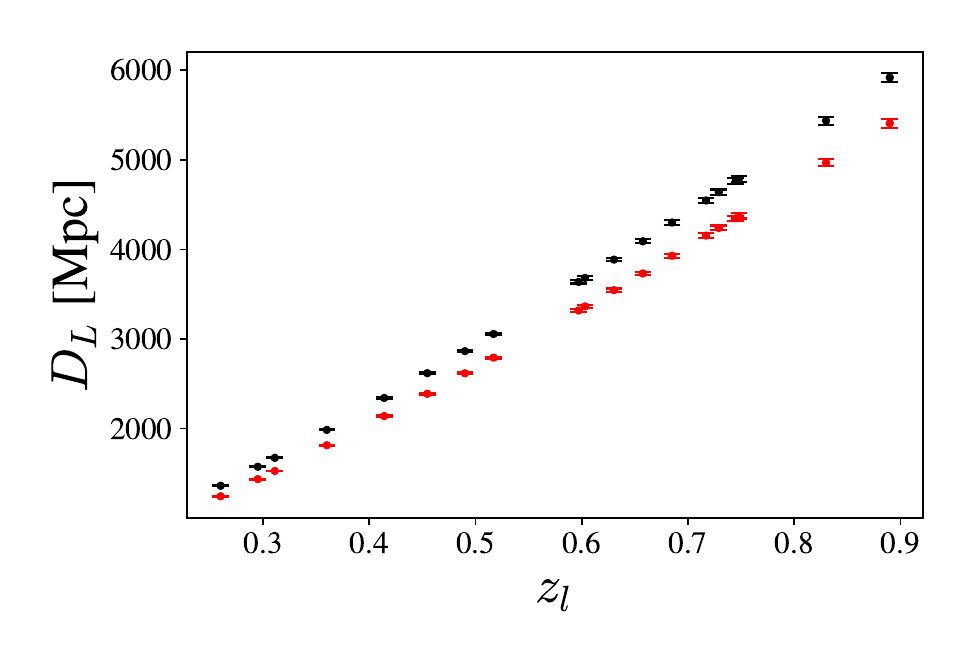} 
\caption{The red points correspond to the $D_L$ data obtained by GP considering  $M_b$  R19, compatible with $H_0$ from the Ref. \cite{Riess2019}.The black points correspond to the $D_L$ data obtained by GP considering  $M_b$ P18, compatible with $H_0$ from the Ref. \cite{Planck2019}. }
\label{Dl:fig}
\end{figure}

The samples of luminosity distance to the lens is obtained from the Pantheon catalog \cite{pantheon}. This is the most recent wide and refined sample of SNe Ia measurements composed by 1048 spectroscopically confirmed SNe Ia and covers a redshift range of $0.01 \leq z \leq 2.3$. We must obtain the SNe Ia at the same redshift to the lens, for this purpose we apply the GP method to find out the central value with the corresponding variance. The first $D_L$ sample is constructed from the apparent magnitude ($m_b$) Pantheon catalog considering the absolute magnitude $M_b = -19.23 \pm 0.04$  (henceforth R19) via the relation:
\begin{equation}
D_L=10^{(m_b-M_b-25)/5} \text{ Mpc}.
\end{equation}This value is obtained by constraining cosmological parameters in a $\Lambda$CDM framework assuming the Hubble rate provided by the Cepheids/SNe Ia estimates, $H_0 = 74.03 \pm 1.42$ km/s/Mpc  \cite{Riess2019}.  The second sample is constructed by using $M_b = -19.43 \pm 0.02$ (henceforth P18) obtained by considering the Hubble rate estimated by the Planck Collaborations in the context of a $\Lambda$CDM model, $H_0 = 67.36 \pm 0.54$ km/s/Mpc \cite{Planck2019}  (see Fig. \ref{Dl:fig}).

\subsection{Time-Delay}

\begin{figure}[htb!]
\centering	
\includegraphics[scale=0.55]{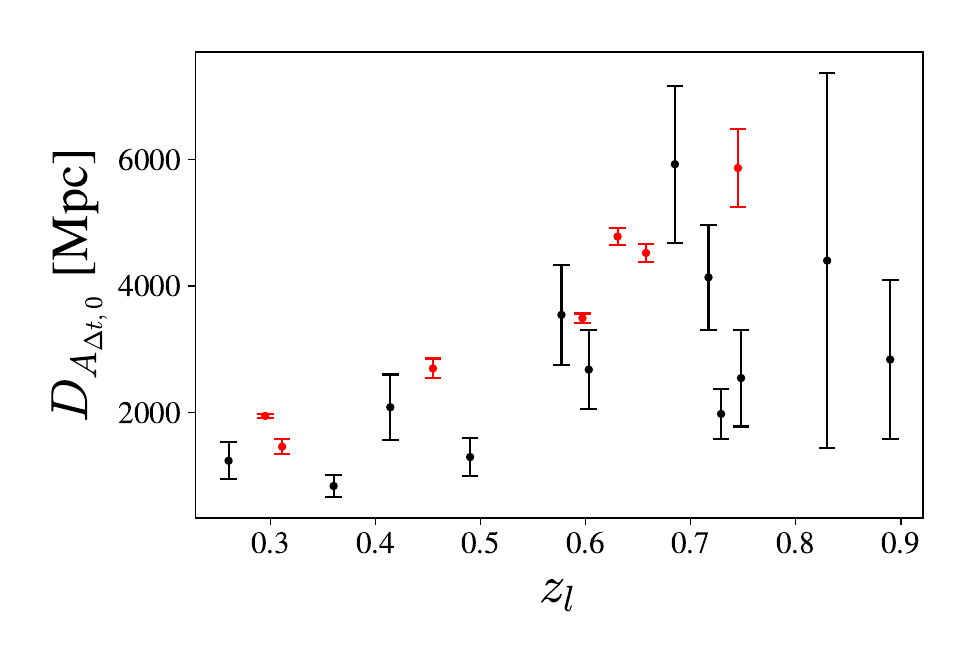} 
\caption{Time-delay angular diameter distance measurements of strong gravitational lensing systems compiled by \cite{ibpsc} (black points) and \cite{holicow} (red points).}
\label{Da:fig}
\end{figure}

We use a data set of 12 two-image time-delay lensing systems compiled by \cite{ibpsc}. The use of two-image lensing systems is justified by the consistency with SIS mass profile and its simplicity. However, this selection criterion is necessary but not sufficient to guarantee a SIS mass profile for the lens. Thus, as mentioned in Ref. \cite{rana2017}, we  include an additional error source denoted by $\zeta$ which takes into consideration possible scatters of individual lenses from a pure SIS mass profile\footnote{Such as the presence of softened isothermal sphere potential, and systematic errors in the RMS deviation of the velocity dispersion.}. Moreover, according to Ref. \cite{scao2012} $\zeta$ can contribute up to $20\%$ in the $D_{A_{\Delta t}}$ estimation. Adding $\zeta$ quadratically the associated error we obtain

\begin{eqnarray}
    \sigma_{D_{A_{\Delta t,0}}}^{2} &=& D_{A_{\Delta t,0}}^2 \left\{ \Bigg( \frac{\sigma_{\Delta t}}{\Delta t}  \Bigg)^2 + 4\left[ \frac{\sigma_{\theta_A}\theta_A}{(\theta_{B}^{2}-\theta_{A}^{2})} \right]^2 \right. \nonumber \\
     && \left. + 4\left[ \frac{\sigma_{\theta_B}\theta_B}{(\theta_{B}^{2}-\theta_{A}^{2})} \right]^2 + \zeta^2 \right\},
\end{eqnarray}
where $\sigma_{\theta_A}$ and $\sigma_{\theta_B}$ are the errors associated with the source images positions A and B, respectively, and $\sigma_{\Delta t}$ the time-delay error. In addition, we consider {  seven more} time-delay systems obtained by the COSMOGRAIL’s Wellspring (H0LiCOW) collaboration and listed in Table 2 of \cite{holicow}. Each system were modeled {  using} constraints from high-resolution HST and/or ground-based AO (Adaptive Optics) imaging data (see more details in \cite{holicow2}). Since  the data from the Ref. \cite{holicow}   present asymmetric error bars, the
data were treated using the method from the Ref. \cite{dag} (see Fig. \ref{Da:fig}).

\subsection{Einstein Radius}
\label{ER}

We consider a specific catalog containing 158 confirmed sources of strong gravitational lensing presented in Ref. \cite{Leaf2018lfu}. This compilation includes 118 SGL systems identical to the compilation in Ref. \cite{Cao2015} along with 40 new systems recently discovered by SLACS and pre-selected by \cite{Shu2017} (see Table I in Ref. \cite{Leaf2018lfu}).

However, studies using SGL systems have shown that the pure SIS model may not be an accurate representation of the lens mass distribution when $\sigma_0<250$ km/s, for which non-physical values of the quantity $D_0$ are usually obtained ($D_0>1$). In addition, in Ref. \cite{Leaf2018lfu} it is mentioned the need for caution when using the SIS model as a reference model, since the impact caused on the density profile can lead to deviations in the observed stellar velocity dispersion ($\sigma_0$). {  It was also observed  the need of introducing an additional  intrinsic error of approximately $12.22\%$ in order to obtain a better concordance between the data and the  $\omega$CDM and $\Lambda$CDM models (see \cite{Leaf2018lfu} for more details)}. 

\begin{figure}[htb!]
\centering	
\includegraphics[scale=0.30]{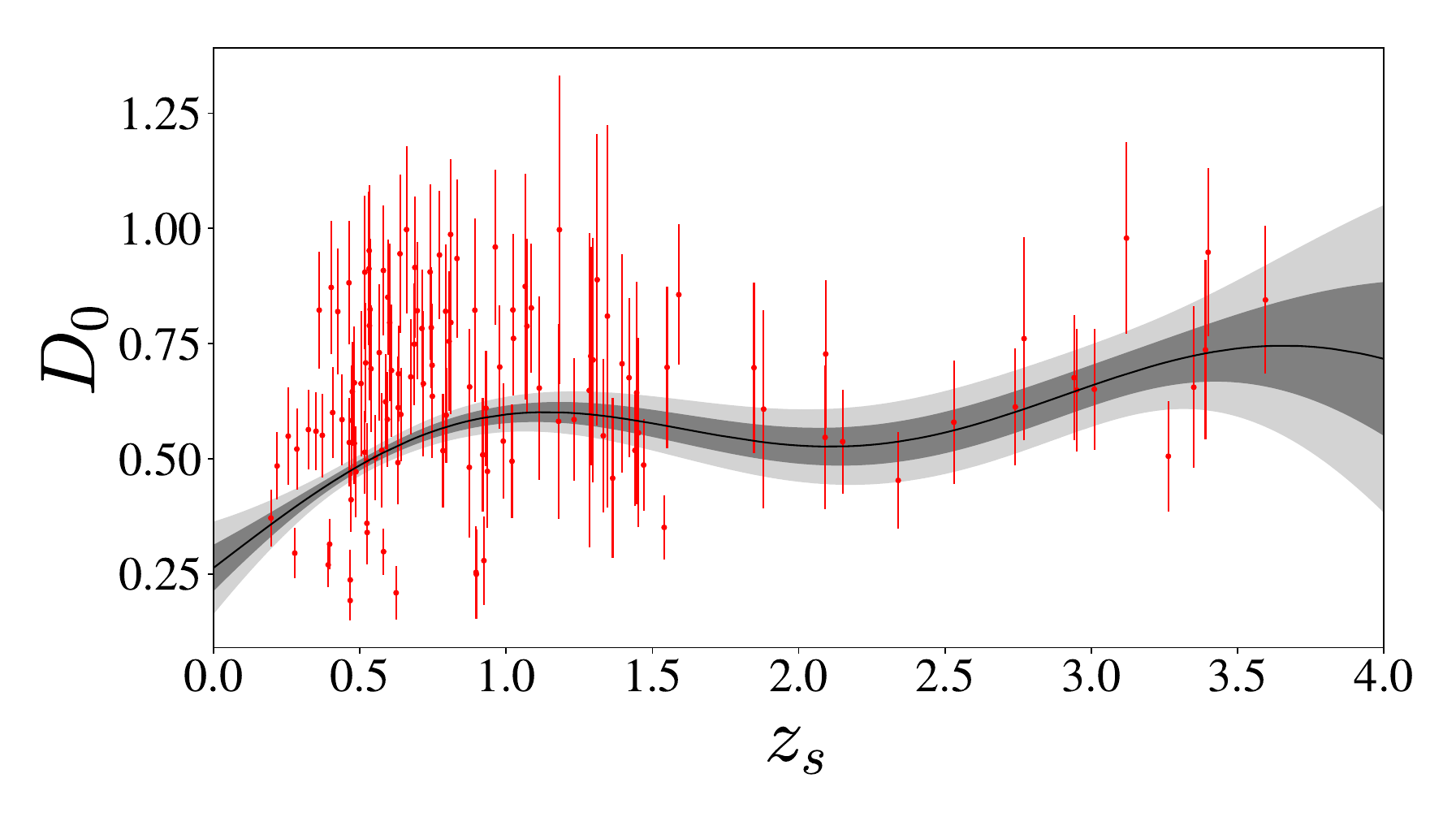} 
\caption{The figure represent an estimate of the ratio $D_0$ for PLAW model. The solid black lines represent the central value reconstruction obtained by the Gaussian Process method with the corresponding $1\sigma$ and $2\sigma$ confidence regions.}
\label{D0:fig}
\end{figure}

Thus, we will consider a general approach to describe the mass distribution of lens-type galaxies, the one in favor of the $\Upsilon$ power-law index model (PLAW), where $\rho \propto r^{- \Upsilon}$. This type of model is important due to several recent studies have shown that the loops of the density profiles of individual galaxies have exhibited a non-negligible spread of the SIS model \cite{1992grlebookS}. Thus, the term $D_0$ of equation (10) is written by:

\begin{equation}
    D_0 = \frac{e^4\theta_E}{4\pi \alpha_{0}^{2}\hbar^2\sigma_{ap}^{2}}f(\theta_E,\theta_{ap}, \Upsilon),
    \label{D0}
\end{equation}
where $ f(\theta_E, \theta_{ap}, \Upsilon)$ is a function which depends on Einstein's radius ($ \theta_E $), the angular aperture used by certain gravitational lens Surveys ($\theta_{ap}$), and the power-law index ($\Upsilon$). In the limit $ \Upsilon=2 $ the SIS model is recovered. Moreover, for a single system we could use the line-of-sight velocity dispersion ($ \sigma_{ap}^{2}$), but as we deal with a sample we must transform all the velocity dispersions measured within an aperture into velocity dispersions within circular aperture of radius ($ R_{\mathbf{eff}}/2$) following the description \cite{1995MNRAS2761341J}: $ \sigma_0 = \sigma_{ap}(\theta_{\mathbf{eff}/(2\theta_{ap})})^{-0.04}$, where $\theta_{\mathbf{eff}}$ is the effective angular radius. In principle, the use of $\sigma_{ap}$ satisfies the model, but the use of $\sigma_0 $ makes the observable $D_0$ more homogeneous for the set of lens located at different redshifts. For that purpose, we just replace $\sigma_{ap}$ for $\sigma_0$ in Eq. (\ref{D0}) \cite{Cao2015} and, therefore, the corresponding error is given by:

\begin{eqnarray}
    \sigma_{D_0}^{2} &=& D_{0}^{2} \left\{ 4\Bigg( \frac{\sigma_{\sigma_{0}}}{\sigma_{0}}  \Bigg)^2 + (1 - \Upsilon)^2\Bigg(  \frac{\sigma_{\theta_E}}{\theta_E} \Bigg)^2 + {\zeta'}^2 \right\}.
\end{eqnarray}
Here we choose $\Upsilon=2.1$ \cite{Ofek:2003sp}. Finally,{  by excluding systems for which the SIS model does not apply} and the source J0850-0347 as mentioned in Ref. \cite{Leaf2018lfu}, the $D_0$ final sample is composed by 124 measurements. {We apply the GP method to this data set to obtain a reconstructed information of the $D_0$ at the same $z$ of time-delay systems. The result of this regression is presented in Fig. \ref{D0:fig} (with 1$\sigma$ and 2$\sigma$ intervals).}

\section{Analysis and Discussion}

We use Markov Chain Monte Carlo (MCMC) methods to calculate the {  posterior} probability distribution functions (pdf) of the free parameter ($\Vec{\Theta} = \gamma$) \cite{foreman}. Thus, the likelihood distribution function is given by:

\begin{equation}
\mathcal{L} (Data|\Vec{\Theta}) = \prod \frac{1}{\sqrt{2\pi} \sigma_{\mu}} exp \Bigg( -\frac{1}{2} \chi^2   \Bigg),
\label{likelihood}
\end{equation}
where

\begin{equation}
\chi^2 =\sum_i \frac{ \left[ W_i - \frac{\phi^{1/2}(z_{l_i})}{\phi^{3}(z_{s_i})} \right]^2}{\sigma_{T_i}^{2}},
\label{chi2}
\end{equation}

\begin{equation}
    W_i \equiv \frac{D_{L_{l_i}}}{D_{0,i} D_{A_{\Delta t,0i}}(1+z_{l_i})^2},
    \label{Wi}
\end{equation}
and 
\begin{equation}
    \sigma_{T_i}^{2} = \sigma_{W_i}^{2} = \sigma_{D_{0,i}}^{2} + \sigma_{D_{A_{\Delta t,0i}}}^{2} + \sigma_{D_{L_{l_i}}}^{2},
\end{equation}
the associated total error, $\phi(z_s) = 1-\gamma\ln{(1+z_s)}$ and $\phi(z_l)=1-\gamma\ln{(1+z_l)}$ (where $\gamma$ is the physical parameter of the model).  In our analyses, we assume a flat prior as $-1.0 \leq \gamma \leq +1.0$.

    As one may see in Eq. \ref{Wi}, in order to perform our analyses, {it is necessary to obtain the measure of the quantity $D_0$ at the redshift of each time-delay system. Then, we use the reconstruction of $D_0$ obtained with GP method (see Sec. \ref{ER}).} Moreover, in order to obtain a $\chi_{\mathbf{red}}^{2} \approx 1$ {  (see Eq. \ref{chi2})} in our analyses, it was added an additional intrinsic error ($\sigma_{int}$).  We estimate it to be $\approx 33\%$ for \cite{ibpsc} and $\approx 23\%$ for \cite{holicow}. For both samples together, $\sigma_{int}$ was estimated to be $\approx 30\%$. {  The addition of  intrinsic error is necessary due to some possible reasons: i)  the reported errors are possibly underestimated, and (ii) that there is an additional unrecognized systematic effect that has yet to be included in the analysis as, for instance, possible random deviations from power law model. This kind of approach has been used in previous analyses with SGL systems (see \cite{scao2012,Leaf2018lfu}).}

Our results are as follows (in 1$\sigma$ c.l.):
\begin{itemize}
\item By considering $D_L$ measurements from the  Pantheon compilation  with $M_b = -19.23 \pm 0.04$, we obtain: $\gamma = -0.03_{-0.06}^{+0.05}$  for the 12 time-delay systems \cite{ibpsc}, and $-0.17_{-0.07}^{+0.06}$  for the 6 time-delay systems \cite{holicow}. For both samples together:   $\gamma = -0.09_{-0.05}^{+0.04}$. 
\item By considering the $D_L$ measurements from Pantheon compilation with  $M_b = -19.43\pm 0.02$, we obtain: $\gamma = +0.00_{-0.05}^{+0.04}$  for the systems in Ref. \cite{ibpsc}, and $\gamma = -0.13_{-0.06}^{+0.05}$  considering the systems in Ref. \cite{holicow}. For both samples together we obtain  $\gamma = -0.06_{-0.04}^{+0.04}$ (see Table I).
\end{itemize}

\begin{figure}[htb!]
\centering	
\includegraphics[scale=0.95]{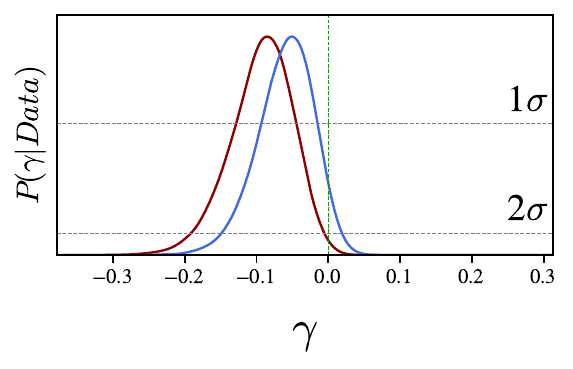}
\caption{The red and blue contours represent the {  posterior} probability distribution functions for $19$ time-delay Systems from \cite{ibpsc,holicow}, but with distinct $D_L$ measurements from Pantheon compilation by \cite{Riess2019} (red) and \cite{Planck2019} (blue). The vertical green dashed line is the limit $\gamma =0$, and the horizontal grey dashed lines represent $1 \sigma$ and $2\sigma$ confidence levels. }
\label{gamma:fig}
\end{figure}

\begin{table}
\small
\centering
	\begin{tabular}{|c|c|c|c|c|} \hline
	\hline
$D_L$ Compilation & time-delay sample   & $\gamma(1\sigma)$ \\ \hline
Pantheon+R19 & \cite{ibpsc} & $-0.03_{-0.06}^{+0.05}$   \\ \hline
Pantheon+P18 & \cite{ibpsc} &  $+0.00_{-0.05}^{+0.04}$  \\ \hline
Pantheon+R19 & \cite{holicow} & $-0.17_{-0.07}^{+0.06} $ \\ \hline
Pantheon+P18 & \cite{holicow} & $-0.13_{-0.06}^{+0.05} $ \\ \hline
Pantheon+R19 & \cite{ibpsc} and \cite{holicow} & $ -0.09_{-0.05}^{+0.04}$ \\ \hline
Pantheon+P18 & \cite{ibpsc} and \cite{holicow} & $-0.06_{-0.04}^{+0.04} $  \\
\hline
\hline    
	\end{tabular}
	\caption{Constraints on a possible time variation of the fine-structure constant for the parameter $\gamma$ of the runaway dilaton model in $1\sigma$ of confidence level.}
    \label{gamma:tab}
\end{table}
{  We also performed our analyses with smoothing technique \cite{Shafieloo,Li,Gonzalez} on $D_0$ quantity, the results are in full agreement with those by using Gaussian Process.}
\section{Conclusions}

The search for a possible time-space variation of the fundamental constants of nature has raised the interest of many cosmologists  due to its possibility of revealing a new underlying physics. In this context, we provided a new method capable of probing a possible temporal evolution of the fine-structure constant by considering time-delay of Strong Gravitational Lensing Systems. A possible temporal variation of $\alpha$ ($\alpha=\alpha_0 \phi(z)$) was investigated in a class of runaway dilaton models, where $\phi(z)=1-\gamma\ln(1+z)$, {  where $\gamma$ is a free parameter}.

The  most restrictive bounds come from the {  joint analysis} using the two samples of time-delay systems together from the Refs. \cite{ibpsc} and \cite{holicow}. We obtained $\gamma = -0.09_{-0.05}^{+0.04}$ and $\gamma = -0.06 _{-0.04}^{+0.04}$ for $D_L$ measurements from Pantheon+R19 and Pantheon+P18, respectively (see Fig. 4 and Table I).

Finally, it is important to stress that recently the authors in the Ref. \cite{Lee:2021kjr} showed that fitting turbulent models in quasars necessarily generates or enhances model non-uniqueness, adding a substantial additional random uncertainty to $\Delta \alpha /  \alpha$ obtained from quasar absorption systems. {  Basically, any given absorption system can be fitted equally well by many slightly different models, each furnishing a different value to $\frac{\Delta \alpha}{\alpha}$.} Therefore, although SGL plus time-delay systems are not as much competitive as the limits imposed by quasar absorption systems, the constraints imposed in this paper provide new and independent limits on a possible temporal variation of the fine-structure constant.

\section{Acknowledgments}

The authors thank to Brazilian scientific and financial support federal agencies, CAPES, and CNPq. RFLH thanks to CNPq No.428755/2018-6 and 305930/2017-6.

\label{lastpage}

\begin{thebibliography}{99}


\bibitem{Dirac:1938mt}{P. A. M. Dirac,  ``\textit{The cosmological constants}'', Nature {\bf 139}, 323 (1937).} 

\bibitem{Uzan:2010pm}{J.~P.~Uzan, ``\textit{Varying Constants, Gravitation and Cosmology}'', Living Rev.\ Rel.\  {\bf 14}, 2 (2011) [arXiv:1009.5514].}

\bibitem{Martins:2017yxk}{C.~J.~A.~P.~Martins, ``\textit{The status of varying constants: a review of the physics, searches and implications}'',  [arXiv:1709.02923].}

\bibitem{Ray:2019lxv}{S. Ray, U. Mukhopadhyay, S. Ray, \& A. Bhattacharjee, ``\textit{Dirac's large number hypothesis: A journey from concept to implication}'', Int.\ J.\ Mod.\ Phys.\ D {\bf 28}, 08 (2019).}

\bibitem{Bekenstein256}{J. D. Bekenstein, ``\textit{Fine-structure constant: Is it really a constant?}'', PRD {\bf 25}, 6 (1982).}

\bibitem{SBM}{H. B. Sandvik, J. D. Barrow, \& J. Magueijo, ``\textit{A simple cosmology with a varying fine-structure constant}'', PRL {\bf 88}, 3 (2002) [astro-ph/0107512].}

\bibitem{BL}{J. D. Barrow, \& S.Z.W. Lip, ``\textit{A Generalized Theory of Varying Alpha}'', PRD {\bf 85}, 023514 (2012) [arXiv:1110.3120].}

\bibitem{BG}{J. D. Barrow, \& A. A. H. Graham, ``\textit{General Dynamics of Varying-Alpha Universes}'', PRD {\bf 88}, 10 (2013) [arXiv:1307.6816].}

\bibitem{MWMK}{M. Dine, W. Fischler, \& M. Srednicki, ``\textit{A simple solution to the strong CP problem with a harmless axion}'', PLB {\bf 104}, 3 (1981); D. B. Kaplan, ``\textit{Opening the axion window}'', NPB {\bf 260}, 1 (1985).}

\bibitem{BBCMAA}{P. Brax, et al., ``\textit{Detecting dark energy in orbit: The cosmological chameleon}'', PRF {\bf 70}, (2004) [arXiv:astro-ph/0408415v2]; P. Brax, C. Van de Bruck, \& A. C. Davies, ``\textit{Compatibility of the ChameleonField Model with Fifth-Force Experiments, Cosmology, and PVLAS and CAST Results}'', PRL {\bf 99}, 12 (2007) [arXiv:hep-ph/0703243v2];  M. Ahlers, A. Lindner, A. Ringwald, L. Schrempp, \& C. Weniger, ``\textit{Alpenglow: A signature for chameleons in axionlike particle search experiments}'', PRD {\bf 77}, 1 (2008) [arXiv:0710.1555v1]; J. Khoury, \& A. Weltman, ``\textit{Chameleon Cosmology}'', PRD {\bf 69}, 4 (2004) [arXiv:astro-ph/0309411v2].}

\bibitem{Chodos}{A. Chodos, \& S.L. Detweiler, ``\textit{Where has the fifth dimension gone?}'', PRD {\bf 21}, 8 (1980).}

\bibitem{wuwang}{Y.S. Wu, \& Z. Wang, ``\textit{Essay on gravitation: Present-time variation of Newton's gravitational constant in superstring theories}'', PRL{\bf 20}, 1 (1988).}

\bibitem{Kiritsis}{E. Kiritsis, ``\textit{Supergravity, D-brane probes and thermal superYang-Mills: A Comparison}'', JHEP {\bf 10}, 010 (1999) [arXiv:hep-th/9906206].}

\bibitem{Landau:2020vkr}{S. J. Landau, ``\textit{Variation of fundamental constants and white dwarfs}'', (2020) [arXiv:2002.00095].}

\bibitem{Bainbridge:2017lsj}{M. B Bainbridge, and others, ``\textit{Probing the Gravitational Dependence of the Fine-Structure Constant from Observations of White Dwarf Stars}'', Universe {\bf 3}, 2 (2017) [arXiv:1702.01757].}

\bibitem{Wilczynska}{M. R. Wilczynska, and others, ``\textit{Four direct measurements of the fine-structure constant 13 billion years ago}'', (2020) [arXiv:2003.07627].}

\bibitem{webb1999}{J. K. Webb, and others, ``\textit{Search for Time Variation of the Fine-Structure Constant}'', PRL {\bf 82}, 5 [astro-ph/9803165].}

\bibitem{Ubachs:2017zmg}{W. Ubachs, ``\textit{Search for varying constants of nature from astronomical observation of molecules}'', Space Sce.\ Rev.\ {\bf 214}, 1 (2018) [arXiv:1709.07704].}

\bibitem{Lee:2021kjr}{C.-C. Lee, J. K. Webb, D. Milakovi\'c, R. F. Carswell, ``\textit{Non-uniqueness in quasar absorption models and implications for measurements of the fine-structure constant}'', (2021) [arXiv:2102.11648].}

\bibitem{Hart:2019dxi}{L. Hart, \& J. Chluba, ``\textit{Updated fundamental constant constraints from Planck 2018 data and possible relations to the Hubble tension}'', MNRAS {\bf 493}, 3 (2020) [arXiv:1912.03986].}

\bibitem{Aghanim:2019ame}{Aghanim, N. and others, ``\textit{Planck 2018 results. V. CMB power spectra and likelihoods}'', A\&A {\bf 641}, A5 (2020) [arXiv:1907.12875].}

\bibitem{Aghanim:2018oex}{Aghanim, N. and others, ``\textit{Planck 2018 results. VIII. Gravitational lensing}'', A\&A {\bf 641}, A8 (2020) [arXiv:1807.06210].}

\bibitem{Ade:2014zfo}{Ade, P.A.R. and others, ``\textit{Planck intermediate results - XXIV. Constraints on variations in fundamental constants}'', A\&A {\bf 580}, A22 (2015) [arXiv:1406.7482].}

\bibitem{Smith:2018rnu}{T. L. Smith, D. Grin, D. Robinson, \& D. Qi, ``\textit{Probing spatial variation of the fine-structure constant using the CMB}'', PRD {\bf 99}, 4 (2018) [arXiv:1808.07486].}

\bibitem{Hees:2020gda}{ A. Hess, and others, ``\textit{Search for a Variation of the Fine-Structure Constant around the Supermassive Black Hole in Our Galactic Center}'', PRL {\bf 124}, 8 (2020) [arXiv:2002.11567].}

\bibitem{galli}{S. Galli, ``\textit{Clusters of galaxues and variation of the fine-structure constant}'', PRD {\bf 87}, 12 (2013) [arXiv:1212.1075v1].}

\bibitem{NBB}{M. T. Clara, \& C. J. A. P. Martins, ``\textit{Primordial nucleosynthesis with varying fundamental constants: Improved constraints and a possible solution to the Lithium problem}'', A\&A {\bf 633}, L11 (2020), [arXiv:2001.01787].}

\bibitem{Milakovic:2020tvq}{D. Milakovi\'c, C.-C. Lee, R. F. Carswell, J. K. Webb, P. Molaro, \& L. Pasquini, ``\textit{A new era of fine-structure constant measurements at high redshift}'', (2020) [arXiv:2008.10619].}

\bibitem{Kraiselburd:2018uac}{L. Kraiselburd, F. L. Castillo, M. E. Mosquera, \& H. Vucetich, ``\textit{Magnetic contributions in Bekenstein type models}'', PRD {\bf 97}, 4 (2018) [arXiv:1801.08594].}

\bibitem{ZhangGengYin}{J.-J. Zhang, L. Yin, \& C.-Q. Geng, ``\textit{Cosmological constraints on $\Lambda(\alpha)$CDM models with time-varying fine-structure constant}'', Annals Phys. {\bf 397}, 400--409 (2018)  [arXiv:1809.04218].}

\bibitem{Wei2017}{H. Wein, X.-B. Zou, H.Y. Li, \& D.Z. Xue, ``\textit{Cosmological constant, fine-structure constant and beyond}'', Eur. Phys. J. C {\bf 77}, 1 (2017) [arXiv:1605.04571].}

\bibitem{Hinkley2013}{N. Hinkley, J. A. Sherman, N. B. Phillips, M. Schioppo, N. D. Lemke, K. Beloy, M. Pizzocaro, C. W. Oates, \& A. D. Ludlow, ``\textit{An Atomic Clock with $10{-18}$ Instability}'', Science {\bf 341}, 6151 (2013) [arXiv:1305.5869].}

\bibitem{Dijck:2020kfb}{E. A. Dijck, ``\textit{Spectroscopy of Trapped $^{138}$Ba$^+$ Ions for Atomic Parity Violation and Optical Clocks}'', (2020).}

\bibitem{damour1}{T. Damour, F. Piazza, \& G. Veneziano, ``\textit{Violations of the equivalence principle in a dilaton-runaway scenario}'', PRD {\bf 66}, 4 (2002) [arXiv:hep-th/0205111v2].}

\bibitem{damour2}{T. Damour, F. Piazza, \& G. Veneziano, ``\textit{Runaway Dilaton and Equivalence Principle Violations}'', PRL  {\bf 89}, 8 (2002) [arXiv:gr-qc/0204094v2].}

\bibitem{colaco2017}{R. F. L. Holanda, L. R. Cola{\c c}o, R. S. Gonalves, \& J. S. Alcaniz, ``\textit{Limits on evolution of the fine-structure constant in runaway dilaton models from Sunyaev-Zeldovich Observation}'', PLB {\bf 767}, 188-192 (2017) [arXiv:1701.07250].}

\bibitem{colaco2019}{L. R. Cola{\c c}o, R. F. L. Holanda, R. Silva, \& J. S. Alcaniz, ``\textit{Galaxy clusters and a possible variation of the fine-structure constant}'', JCAP {\bf 03}, 014 (2019) [arXiv:1901.10947].}

\bibitem{kbora}{K. Bora, \& S. Desai, ``\textit{Constraints on variation of the fine-structure constant from joint SPT-SZ and XMM-Newton observations}'', (2020) [arXiv:2008.10541].}

\bibitem{colaco2020}{L. R. Cola{\c c}o, R. F. L. Holanda, \& R. Silva, ``\textit{Probing variation of the fine-structure constant using the strong gravitational lensing}'', (2020) [arXiv:2004.08484].}

\bibitem{holanda2016.1}{R. F. L. Holanda, S. J. Landau, J. S. Alcaniz, I. E. Sanchez, \& V. C. Busti, ``\textit{Constraints on a possible variation of the fine-structure constant from galaxy cluster data}'', JCAP {\bf 1605}, 047 (2016) [arXiv:1510.07240].}

\bibitem{holanda2016.2}{R. F. L. Holanda, V. C. Busti, L. R. Cola{\c c}o, J. S. Alcaniz, \& S. J. Landau, ``Galaxy clusters, type Ia supernovae and the fine-structure constant'', JCAP {\bf 1608}, 055 (2016) [arXiv:1605.02578].}

\bibitem{ibpsc}{I. Balm{\`e}s, \& P. S. Corasaniti, ``\textit{Bayesian approach to gravitational lens model selection: constraining H$_{0}$ with a selected sample of strong lenses}'', MNRAS {\bf 431}, 2 (2013) [arXiv:1206.5801].}

\bibitem{holicow}{S. Birrer, et al., ``\textit{TDCOSMO - IV. Hierarchical time-delay cosmography – joint inference of the Hubble constant and galaxy density profiles}'', Astron. Astrophys. {\bf 643}, A165 (2020) [arXiv:2007.02941].}

\bibitem{pantheon}{D. M. Scolnic, \textit{et al}., ``\textit{The complete Ligh-curve Sample of Spectroscopically Confirmed SNe Ia from Pan-STARRS1 and Cosmological Constraints from the Combined Pantheon Sample}'', ApJ {\bf 859}, 101 (2018) [arXiv:1710.00845].} 
\bibitem{Leaf2018lfu}{K. Leaf, \& F. Melia, ``\textit{Model selection with strong-lensing systems}'', MNRAS {\bf 478}, 4 (2018) [arXiv:1805.08640].}

\bibitem{Observables}{A. Hess, O. Minazzoli, \& J. Larena, ``\textit{Observables in theories with a varying fine-structure constant}'', Gen. Rel. Grav. {\bf 47}, 2 (2015) [arXiv:1409.7273].}

\bibitem{hees}{O. Hees, A. Minazzoli, \& J. Larena, ``\textit{Breaking of the equivalence principle in the electromagnetic sector and its cosmological signatures}'', PRD {\bf 90}, 12 (2014) [arXiv:1406.6187v4].}

\bibitem{Cao2015}{S. Cao, M. Biesiada, \& R. Gavazzi, ``\textit{Cosmology with Strong-Lensing Systems}'', ApJ {\bf 806}, 2 (2015) [arXiv:1509.07649].}

\bibitem{sef}{P. Schneiner, J. Ehlers, \& E. E. Falco, ``\textit{Gravitational Lenses}'', Springer-Verlag Berlin Heidelberg New York. Also Astronomy and Astrophysics Library 2019.}
  
\bibitem{Refsdal}{S. Refsdal, ``\textit{On the possibility of determining Hubble's parameter and the masses of galaxies from the gravitational lens effect}'', MNRAS {\bf 128}, 307 (1964).}


\bibitem{SHSUYU}{S.H. Suyu, et al., ``\textit{Dissecting the Gravitational lens B1608+656. II. Precision Measurements of the Hubble Constant, Spatial Curvature, and the Dark Energy Equation of State}'', ApJ {\bf 711}, 1 (2009) [arXiv:0910.2773].}

\bibitem{TTREU}{T. Treu, ``\textit{Strong Lensing by Galaxies}'', ARAA {\bf 48}, 87-125 (2010) [arXiv:1003.5567].}
  
 \bibitem{Suyu}S. H. Suyu et al., ``\textit{Dissecting the Gravitational lens B1608+656. II. Precision Measurements of the Hubble Constant, Spatial Curvature, and the Dark Energy Equation of State}'', ApJ {\bf 711}, 201-125 (2010).
 
\bibitem{JLWEI}{J.-L. Wei, X.-F. Wu, \& F. Melia, ``\textit{A Comparison of Cosmological Models Using Time Delay Lenses}'', ApJ {\bf 788}, 190 (2014) [arXiv:1405.2388].}



\bibitem{Minazzoli}{O. Minazzoli, \& A. Hees, ``\textit{Late-time cosmology of a scalar-tensor theory with a universal multiplicative coupling between the scalar field and the matter Lagrangian}'', PRD {\bf 90}, 2 (2014) [arXiv:1404.4266v2].}
 
\bibitem{bxuqhuang}{B. Xu, and Q. Huang, ``\textit{New tests of the cosmic distance duality relation with the baryon acoustic osculattion and type Ia supernovae}'', The Eur. Phys. J. P. {\bf 135}, 06 (2020).}

\bibitem{cmapscorasaniti}{C. Ma, \& P.-S. Corasaniti, ``\textit{Statistical Test of Distance-Duality Relation with Type Ia Supernovae and Baryon Acoustic Oscillatuion}'', ApJ {\bf 861}, 2 (2018) [arXiv:1604.04631].}
 


\bibitem{Seikel:2013fda}{M. Seikel, C. Clarkson, ``\textit{Optimising Gaussian processes for reconstructing dark energy dynamics from supernovae}'', (2013) [arXiv:1311.6678].}

\bibitem{Riess2019}{A. G. Riess, S. Casertano, W. Yuan, L. M. Macri, D. Scolnic, ``\textit{Large Magellanic Cloud Cepheid Standards Provide a 1\% Foundation for the Determination of the Hubble Constant and Stronger Evidence for Physics beyond $\Lambda$CDM}'', Astrophys. J. {\bf 876}, 1 (2019) [arXiv:1903.07603].}

\bibitem{Planck2019}{N. Aghanim, and others, ``\textit{Planck 2018 results. VI. Cosmological parameters}'', Astron. Astrophys. {\bf 641}, A6 (2020) [arXiv:1807.06209].} 

\bibitem{rana2017}{A. Rana, D. Jain, S. Mahajan, A. Mukherjee, \& R. F. L. Holanda, ``\textit{Probing the cosmic distance duality relation using time delay lenses}'', JCAP {\bf 07}, 010 (2017) [arXiv:1705.04549].}

\bibitem{scao2012}{S. Cao, Y. Pan, M. Biesiada, W. Godlowski, \& Z.-H. Zhu, ``\textit{Constraints on cosmological models from strong gravitational lensing systems}'', JCAP {\bf 2012}, 3 (2012) [arXiv:1105.6226].}

\bibitem{holicow2}{S. Birrer, and others, ``\textit{TDCOSMO - IV. Hierarchical time-delay cosmography \textendash{} joint inference of the Hubble constant and galaxy density profiles}'', Astron. Astrophys. {\bf 643}, A165 (2020) [arXiv:2007.02941]; K. C. Wong, and others, ``\textit{H0LiCOW \textendash{} XIII. A 2.4 per cent measurement of H0 from lensed quasars: 5.3\ensuremath{\sigma} tension between early- and late-Universe probes}'', Mon. Not. Roy. Astron. Soc. {\bf 498}, 1 (2020) [1907.04869].}

\bibitem{dag}{D’Agostini, G., ``\textit{Asymetric uncertainties: Sources, treatment and potential dangers}'', (2004) [physics/0403086].}

\bibitem{Shu2017}{Y. Shu, et al., ``\textit{The Sloan Lens {ACS} Survey. {XIII}. Discovery of 40 New Galaxy-scale Strong Lenses}'', ApJ {\bf 851}, 1 (2017) [arXiv:1711.00072].}

\bibitem{1992grlebookS}{P. Schneider, J. Ehlers, \& E. E. Falco, ``\textit{Gravitational Lenses}'', Springer Science \& Business Media (1999).}

\bibitem{1995MNRAS2761341J}{I. Jorgensen, M. Franx, \& P. Kjaergaard, ``\textit{Spectroscopy for E and S0 galaxies in nine clusters}'', MNRAS {\bf 276}, 4 (1995).}

\bibitem{Ofek:2003sp}{E. O. Ofek, H.-W. Rix, \& D. Maoz, ``\textit{The redshift distribution of gravitational lenses revisited: Constraints on galaxy mass evolution}'', MNRAS {\bf 343}, 639 (2003) [astro-ph/0305201].}

\bibitem{foreman}{D. Foreman-Mackey, D. W. Hogg, D. Lang, \& J. Goodman, ``\textit{emcee: The MCMC Hammer}'', Publ.\ Astron.\ Soc.\ Pac.\ {\bf 125}, 925 (2013) [arXiv:1202.3665].}

\bibitem{Shafieloo}{A. Shafieloo, U. Alam, V. Sahni, \& A. Starobinsky, ``\textit{Smoothing Supernova Data to Reconstruct the Expansion History of the Universe and its Age}'', MNRAS {\bf 366}, 1081 (2006)[arXiv:0505329].}

\bibitem{Li}{Li, Zhengxiang and Gonzalez, J. E. and Yu, Hongwei and Zhu, Zong-Hong and Alcaniz, J. S.",
   ``\textit{Constructing a cosmological model-independent Hubble diagram of type Ia supernovae with cosmic chronometers}",
    Phys. Rev. D {\bf 93}, 4 (2016) [arXiv:1504.03269].
}

\bibitem{Gonzalez}{Gonzalez, J. E. and Alcaniz, J. S. and Carvalho, J. C.,'' \textit{Smoothing expansion rate data to reconstruct cosmological matter perturbations}'',  JCAP {\bf 08}, 008  (2017)
[arXiv:1702.02923].} 
\end{thebibliography}
\end{document}